\documentclass[prd,aps,floats,onecolumn,twoside,preprintnumbers,
superscriptaddress,floatfix,nofootinbib,showpacs]{revtex4}
\usepackage{graphicx,pstricks,rotating,amsmath,epstopdf}
\usepackage{amssymb,bbm}
\usepackage{slashed}
\begin{document}   


\renewcommand{\Im}{\mathop{\rm Im}}
\setcounter{tocdepth}{3}

\title{\center{Nonassociative Snyder $\phi^4$ Quantum Field Theory}}

\author{Stjepan Meljanac}
\email{meljanac@irb.hr}
\affiliation{Rudjer Bo\v skovi\' c Institute, Division of Theoretical Physics, P.O.Box 180, HR-10002 Zagreb, Croatia}
\email{meljanac@irb.hr}
\author{Salvatore Mignemi}
\affiliation{Dipartimento di Matematica e Informatica, Universita di Cagliari,\\
viale Merello 92, 091123 Cagliari, Italy}
\email{smignemi@unica.it}
\affiliation{INFN, Sezione di Cagliari, 09042 Monserrato, Italy}
\author{Josip Trampetic}
\affiliation{Rudjer Bo\v skovi\' c Institute,  Division of Experimental Physics, P.O.Box 180, HR-10002 Zagreb, Croatia}
\email{josip.trampetic@irb.hr}
\affiliation{Max-Planck-Institut f\"ur Physik, (Werner-Heisenberg-Institut),
\\F\"ohringer Ring 6, D-80805 M\"unchen, Germany}
\email{trampeti@mppmu.mpg.de}
\author{Jiangyang You}
\email{youjiangyang@gmail.com}
\affiliation{Rudjer Bo\v skovi\' c Institute, Theoretical Physics Division, P.O.Box 180, HR-10002 Zagreb, Croatia}

\date{\today}

\begin{abstract} In this article we define and quantize a truncated form of the nonassociative and noncommutative Snyder  $\phi^4$ field theory using the functional method in momentum space. More precisely, the action is approximated by expanding up to the linear order in the Snyder deformation parameter $\beta$, producing an effective model on commutative spacetime for the computation of the two-, four- and six-point functions. The two- and four-point functions at one loop have the same structure as at the tree level, with UV divergences faster than in the commutative theory. The same behavior appears in the six-point function, with a logarithmic UV divergence and renders the theory unrenormalizable at $\beta^1$ order except for the special choice of free parameters $s_1=-s_2$. We expect effects from nonassociativity on the correlation functions at $\beta^1$ order, but these are cancelled due to the average over permutations.
\end{abstract}



\maketitle

\section{Introduction}

There is consensus in the theoretical and mathematical physics
nowadays  that at short distances spacetime has to be described
by nonstandard geometrical structures, and that the very concept of
point and localizability may no longer be adequate. Together with string
theories \cite{Seiberg:1999vs}, this is one of the oldest motivations for 
the introduction of noncommutative (NC) geometry~
\cite{Connes:1994yd,Doplicher:1994tu,Majid:1996kd,Landi:1997sh,Madore:1999bi,Madore:2000aq,GraciaBondia:2001tr}. The simplest kind of noncommutative geometry is the so-called ``canonical'' one~\cite{Doplicher:1994tu,Szabo:2001kg,Douglas:2001ba,Moyal:1949sk,Szabo:2009tn,Schupp:2002up,Schupp:2008fs}. Usually, the construction of a field theory on a noncommutative space is performed by deforming the product between functions (and, hence, between fields in general) with the introduction of a noncommutative star product. The noncommutative coordinates $\hat{x}^\mu$ satisfy
\begin{equation}
[\hat{x}^\mu,\hat{x}^\nu]=i \theta^{\mu\nu},
\label{commutatorxMoy}
\end{equation}
with coordinates $x^\mu$ being promoted to Hermitian operators $\hat{x}^\mu$ satisfying (\ref{commutatorxMoy}). Note that the choice of the star($\star$)-product compatible with (\ref{commutatorxMoy}) is not unique.

The simplest case  $|\theta^{\mu\nu}|\sim$ const is the well-known
Moyal noncommutative spacetime \cite{Moyal:1949sk}: $|\theta^{\mu\nu}|$ does not depend on coordinates, and it scales like length$^2\sim \Lambda^{-2}_{\rm NC}$, $\Lambda_{\rm NC}$ being the scale of noncommutativity with the dimension of energy. For Moyal geometry, it was proven recently that there exists a $\theta$-exact formulation of noncommutative gauge field theory based on the Seiberg-Witten map \cite{Seiberg:1999vs,Schupp:2008fs} that preserves unitarity \cite{Aharony:2000gz} and has improved UV/IR behavior at the quantum level by introducing supersymmetry \cite{Horvat:2011bs,Martin:2016zon,Martin:2016hji,Martin:2016saw}. All these could  also have implications for cosmology, for example, through the determination of the maximal decoupling temperature of the right-handed neutrino species in the early Universe \cite{Horvat:2017gfm}.

There are other important models, like the $\kappa$-Minkowski and the Snyder geometries, where we might expect  similar properties with analogous cosmological consequences. For example, results in \cite{Horvat:2017gfm} represent one of the strongest  motivation for our  investigation of Snyder spaces.

The $\kappa$-Minkowski models \cite{Lukierski:1991pn,Lukierski:1992dt,Meljanac:2007xb,Meljanac:2010qp,Govindarajan:2008qa,Govindarajan:2009wt,Meljanac:2011cs}, are represented by 
\begin{equation}
[\hat{x}_\mu,\hat{x}_\nu]=
\frac{i}{\kappa}(\delta_{\mu}^{\;\;0}\;\hat x^{}_{\nu}-\delta_{\nu}^{\;\;0}\;\hat x^{}_{\mu}),
\label{commutatorxMink}
\end{equation}
where $\kappa$ is a mass parameter. On the other hand, Snyder's spacetime \cite{Snyder:1946qz}, the subject of this investigation, belongs to a rather different type of models \cite{Maggiore:1993kv,Battisti:2010sr,Mignemi:2013aua,Mignemi:2015fva}, and is defined by the phase space commutation relations,
\begin{equation}
[\hat{x}_\mu,\hat{x}_\nu]=i\beta M_{\mu\nu},\;
[p^\mu,\hat{x}_\nu]=-i\delta^{\mu}_{\;\;\nu}-i\beta p^{\mu}p_{\nu},\;
[p_\mu,p_\nu]=0,
\label{commutatorxSnyd}
\end{equation}
where $M_{\mu\nu}=x_\mu p_\nu-x_\nu p_\mu$ are Lorentz generators, $x_\mu$ are the undeformed canonical coordinates and $p_\mu$ the momentum generators. Noncommutative coordinates $\hat{x}_\mu$ and momentum generators $p_\mu$ transform as vectors under Lorentz generators and $\beta$ is a real parameter $\beta \propto\ell^2_P$, where $\ell_P$ is the Planck length.

The Moyal and the $\kappa$-Minkowski geometries break the Lorentz invariance. Such effects are manifested in their star product. On the contrary, in his seminal paper Snyder \cite{Snyder:1946qz} observed that assuming a noncommutative structure of spacetime and hence a deformation of the Heisenberg algebra it is possible to define a discrete spacetime without breaking  the Lorentz invariance. It is, therefore, interesting to investigate  the Snyder model from the general point of view of noncommutative geometry.

More recently, the formalism of Hopf algebras has been applied to the study of noncommutative geometries \cite{Majid:1996kd}. The Snyder model has been studied in a series of papers \cite{Battisti:2010sr,Mignemi:2013aua,Mignemi:2015fva,Lu:2011it,Lu:2011fh,Girelli:2010wi,Meljanac:2016gbj} and the associated Hopf algebra investigated in \cite{Battisti:2010sr} and \cite{Meljanac:2016gbj}, where the model has been generalized and the star product, coproducts and antipodes have been calculated using the method of realizations.
A different approach was used in \cite{Girelli:2010wi},  where the Snyder model was considered in a geometrical perspective as a coset in momentum space, and the results are equivalent to those of Refs.  \cite{Mignemi:2013aua,Mignemi:2015fva}.  A further generalization of Snyder spacetime deformations was recently introduced in \cite{Meljanac:2016gbj,Meljanac:2016jwk,Meljanac:2017ikx}. Also several nonassociative star/cross product geometries and related quantum field theories have been discussed recently in \cite{Kupriyanov:2017oob}.

In this paper we consider a Snyder-like quantum field theory, where the action is modified by truncating the model to first order in the deformation parameter $\beta$. The drawback of this truncation is the loss of the
ultraviolet behavior of the original theory. In particular, we remark that the original theory could be ultraviolet finite. Moreover, any possible nonperturbative effect like the celebrated UV/IR mixing in \cite{Minwalla:1999px,Grosse:2005iz,Schupp:2008fs} is also lost. Among other features, UV/IR mixing connects the noncommutative field theories  with holography via UV and IR cutoffs   in a model independent way \cite{Cohen:1998zx,Horvat:2010km}. Holography and UV/IR mixing are known in the literature as possible windows to quantum gravity \cite{Cohen:1998zx,Szabo:2009tn}. In spite of this deficiency, we believe that our investigation is interesting as a starting point for further investigations on the properties of the full theory.

The paper is organized as follows: in the second section, we introduce  the Hermitian realization of the model and the star product  corresponding to this realization. The  Snyder-deformed action for a $\phi^4$ theory based on the above formalism is introduced in Sec. III. The quantization of the theory, including the tree-level, four-point function, as well as the one-loop two-, four-, and six-point functions, is discussed  in Section 4. The effect of Snyder's nonassociativity is presented in Sec. V. Finally, in Sec. VI, we discuss the UV divergences and their possible disappearance in the full theory.

\section{Hermitian realization of Snyder spaces}

Following Refs.~\cite{Meljanac:2016gbj,Meljanac:2017ikx}, we consider the Hermitian realization of the Snyder spaces
\begin{equation}
\hat x_\mu=x_\mu+\beta\left[s_1 M_{\mu\alpha}p^\alpha +(s_1+s_2)(x\cdot p) p_\mu-i\Big(s_1+\frac{D+1}{2}s_2\Big)p_\mu\right]+\mathcal O(\beta^2),
\label{xreal}
\end{equation}
with $D$ the dimension of the spacetime we are considering,\footnote{We write directly $D$ here since this factor later enters the loop computation and we use dimensional regularization when evaluating loop integrals. Dimensional regularization appears to be a natural choice because there is no tensor structure other than metric in our formulation of the Snyder theory and so we only encounter scalar and vector objects and no pseudoscalars or pseudovectors.}  and $s_1,s_2$  real parameters. The generators $M_{\mu\nu},\;  p_\mu$, and $x_\mu,\; p_\mu$, generate the undeformed Poincar\'e and Heisenberg algebras, respectively.
The commutation relations $[\hat{x}_\mu,\hat{x}_\nu],\;[p^\mu, \hat{x}_\nu]$ are
\begin{equation}
[\hat{x}_\mu,\hat{x}_\nu]=i\beta (s_2-2s_1)M_{\mu\nu}+ \mathcal O(\beta^2),\;[p^\mu, \hat{x}_\nu]=-i\Big(\delta^\mu_\nu\big(1+\beta s_1 p^2\big)+\beta s_2p^\mu p_\nu\Big)+ \mathcal O(\beta^2),
\label{commutnew}
\end{equation}
which implies that the coordinates $\hat{x}_\mu$ become commutative for $s_2=2s_1$.

The corresponding star product takes the following form
\begin{equation}
e^{ikx}\star e^{iqx}=e^{iD_\mu(k,q) x^\mu}e^{iG(k,q)},
\label{starproduct}
\end{equation}
and it is in general nonassociative and noncommutative. However, for specific choice $s_2=2s_1$ in (\ref{commutnew}), the star product (\ref{starproduct}) becomes associative and commutative. The functions $D_\mu(k,q)$ and $G(k,q)$ are given up to first order in $\beta$ for arbitrary $s_1$ and $s_2$ by
\begin{eqnarray}
D_\mu(k,q)&=&k_\mu+q_\mu
\label{Dmu}\\
&+&\beta\bigg[k_\mu\bigg(s_1 q^2+\Big(s_1+\frac{s_2}{2}\Big)k\cdot q \bigg)
+q_\mu s_2\Big(k\cdot q+\frac{k^2}{2}\Big)\bigg]+\mathcal O(\beta^2),
\nonumber\\
G(k,q)&=&-i\beta\Big(s_1+\frac{D+1}{2}s_2\Big)k\cdot q+\mathcal O(\beta^2),
\label{G}
\end{eqnarray}
and they satisfy relation
\begin{equation}
\det\left(\frac{\partial D_\mu(k,q)}{\partial k_\nu}\right)\bigg|_{k=-q}=\det\left(\frac{\partial D_\mu(k,q)}{\partial q_\nu}\right)\bigg|_{k=-q}=e^{iG(k,-k)}+\mathcal O(\beta^2),
\label{detD}
\end{equation}
which induces the cyclicity of the star product under usual integration
\begin{equation}
\int f(x)\star g(x)=\int f(x)g(x)+\mathcal O(\beta^2).
\label{cyclicity}
\end{equation}
In other words, a usual integration removes the effects of the deformation  by at least one order, since both $D_\mu(k,q)$ and $G(k,q)$ contain $\mathcal O(\beta^1)$ terms while any deformation effect in \eqref{cyclicity} must start at $\mathcal O(\beta^2)$. Note that, in principle, an integral over any star product of two fields under the Hermitian realization condition would reduce to the integral of the usual multiplication. This is certainly true for Moyal and $\kappa$-Minkowski cases; however, for the above conjecture in the general case of Snyder spaces, we only have a rigorous proof up to the $\mathcal O(\beta^2)$ and in the Snyder realization of the full theory \cite{Meljanac:2017ikx}.

Note also that using Eqs (\ref{starproduct})-(\ref{G}) it is straightforward to show that the 3-cyclicity for nonassociative star products
\begin{equation}
\int f\star(g\star h)=\int (f\star g)\star h,
\label{11}
\end{equation}
is lost except when $s_2=2s_1$.

\section{The $\phi^4$ theory on Snyder spaces}

The action for a Snyder-type $\phi^4$ theory on four-dimensional Euclidean spacetime\footnote{In order to avoid complications we choose to work directly on Euclidean spacetime.} is given by
\begin{equation}
S=\int \frac{1}{2}\Big((\partial_\mu\phi)\star(\partial^\mu\phi)+m^2\phi\star\phi\Big)+S_{int},
\label{phi4}
\end{equation}
where
\begin{equation}
S_{int}=-\frac{\lambda}{4!}\int\phi\star(\phi\star(\phi\star\phi)).
\label{Sint}
\end{equation}
Up to the first order in $\beta$ we can remove the  star product on the left using the cyclicity property of the star product~\eqref{cyclicity} to get
\begin{equation}
S^1=\int \frac{1}{2}\Big((\partial\phi)^2+m^2\phi^2\Big)-\frac{\lambda}{4!}\phi(\phi\star(\phi\star\phi))
+\mathcal O(\beta^2).
\label{S1}
\end{equation}
The definition~\eqref{starproduct} of the  star product then allows us to write the interaction in  momentum space as follows
\begin{equation}
\begin{split}
S^1_{int}=&-\frac{\lambda}{4!}\int\phi(\phi\star(\phi\star\phi))
\\=&-\frac{\lambda}{4!}\int\frac{d^4q_1}{(2\pi)^4}\frac{d^4q_2}{(2\pi)^4}\frac{d^4q_3}{(2\pi)^4}\frac{d^4q_4}{(2\pi)^4}g_3(q_1,q_2,q_3,q_4)
\\&\cdot(2\pi)^4\delta\big(D_4(q_1,q_2,q_3,q_4)\big)\tilde\phi(q_1)\tilde\phi(q_2)\tilde\phi(q_3)\tilde\phi(q_4)
+\mathcal O(\beta^2),
\end{split}
\label{Sint1}
\end{equation}
where
\begin{equation}
D_4(q_1,q_2,q_3,q_4)=q_1+D(q_2,D(q_3,q_4)),
\label{D4}
\end{equation}
and
\begin{equation}
g_3(q_1,q_2,q_3,q_4)=1+iG(q_2,D(q_3,q_4))+iG(q_3,q_4)+\mathcal O(\beta^2).
\label{g3}
\end{equation}
This is our starting point for the following calculations.

\section{Quantizing the Snyder field theory}

Since the quadratic part of the classical action is undeformed, it is convenient to adopt the functional method in momentum space, previously used in similar problems like for example~\cite{Grosse:2005iz}.
Our starting point is the generating functional
\begin{equation}
Z[J]=e^{W[J]}=\exp\left[-S+\int J\phi\right],
\label{ZJ}
\end{equation}
which we shall evaluate perturbatively. The generating functional for the free theory is
\begin{equation}
Z_0[J]=\exp\left[\int d^4x d^4 y J(x)G(x-y)J(y)\right].
\label{Z0J}
\end{equation}
Since the free Euclidean Green's function is simply
\begin{equation}
G(x-y)=\int \frac{d^4 k}{(2\pi)^4}\frac{e^{ik(x-y)}}{k^2+m^2},
\label{EGreen}
\end{equation}
the free generating functional can be reduced to the momentum space expression
\begin{equation}
\begin{split}
Z_0[J]&=\exp\left[\int d^4x d^4 y J(x)G(x-y)J(y)\right]
=\exp\left[\int \frac{d^4 k}{(2\pi)^4} \tilde J(k)\frac{1}{k^2+m^2}\tilde J(-k) \right].
\end{split}
\label{Z0JMS}
\end{equation}

The generating functional of the interacting theory is obtained by introducing the interaction through functional derivatives of the free generating functional, i.e.
\begin{equation}
\begin{split}
Z[J]=&\mathcal N\exp\bigg[\frac{\lambda}{4!}\int\frac{d^4q_1}{(2\pi)^4}\frac{d^4q_2}{(2\pi)^4}\frac{d^4q_3}{(2\pi)^4}\frac{d^4q_4}{(2\pi)^4}g_3(q_1,q_2,q_3,q_4)
\\&\cdot(2\pi)^4\delta\big(D_4(q_1,q_2,q_3,q_4)\big)\frac{\delta}{\delta\tilde J(q_1)}\frac{\delta}{\delta\tilde J(q_2)}\frac{\delta}{\delta\tilde J(q_3)}\frac{\delta}{\delta\tilde J(q_4)}\bigg]Z_0[J].
\end{split}
\label{ZJMS}
\end{equation}
The functional derivative $\frac{\delta}{\delta\tilde J(q)}$ satisfies
\begin{equation}
\frac{\delta}{\delta\tilde J(q)}\tilde J(p)=(2\pi)^4\delta(p-q),
\label{Fd}
\end{equation}
where the factor $(2\pi)^4$ follows from the normalization adopted for the Fourier transformation,
\begin{equation}
\phi(x)=\int \frac{d^4p}{(2\pi)^4}e^{ipx}\tilde\phi(p).
\label{Fourier}
\end{equation}

The Green's function obtained from the generating functional contains, in principle, a number of $\delta$ functions, in particular the composite ones on the vertices, so we need a strategy to handle them properly. We choose the following prescription:  first we work on the position space connected correlation functions
\begin{equation}
\begin{split}
G(x_1,x_2,.....,x_n)=\int\prod\limits_{i=1}^n\frac{d^D p_i}{(2\pi)^D}e^{ip_ix_i}\frac{\delta}{\delta\tilde J(p_i)}W[J]\bigg|_{J=0},
\end{split}
\label{Gx1...n}
\end{equation}
because all external and internal momenta  are integrated over and consequently all $\delta$functions can be evaluated as well. We then integrate over one specific fixed external momentum $p_n$ in order to remove the final (composite) $\delta$ function that describes the modified overall momentum conservation. This is not the only possible choice one could make, but we will stick with it and construct both tree and one-loop level integrals accordingly.

\subsection{Tree-level four-point function}

\begin{figure}[t]
\begin{center}
\includegraphics[width=4cm,angle=0]{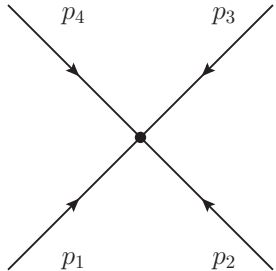}
\end{center}
\caption{Four-point Feynman rule.}
\label{fig:FD1}
\end{figure}

As an example of the method described in the last section as well a basis for the further  computations, we evaluate first the tree-level, four-point correlation function, $G_{tree}(x_1,x_2,x_3,x_4)$ (corresponding  to  Fig.\ref{fig:FD1}), which is defined as follows:
\begin{eqnarray}
G_{tree}(x_1,x_2,x_3,x_4)
&=&\frac{\lambda}{4!}\int\frac{d^4 p_1}{(2\pi)^4}\frac{d^4 p_2}{(2\pi)^4}\frac{d^4 p_3}{(2\pi)^4}\frac{d^4 p_4}{(2\pi)^4}
\frac{e^{ip_1x_1}}{p_1^2+m^2}\frac{e^{ip_2x_2}}{p_2^2+m^2}\frac{e^{ip_3x_3}}{p_3^2+m^2}\frac{e^{ip_4x_4}}{p_4^2+m^2}
\nonumber\\
&\cdot&(2\pi)^4\sum\limits_{\sigma\in S_4}\delta\Big(D_4\big(\sigma(p_1,p_2,p_3,p_4)\big)\Big)\cdot g_3\big(\sigma(p_1,p_2,p_3,p_4)\big),
\label{Gtree}
\end{eqnarray}
where $\sigma\in S_4$ denotes the sum over all  momenta permutations, i.e.
\begin{gather}
\delta\Big(D_4\big(\sigma(p_1,p_2,p_3,p_4)\big)\Big)=\delta\Big(D_4(q_1=p_{\sigma(1)},q_2=p_{\sigma(2)},q_3=p_{\sigma(3)},q_4=p_{\sigma(4)})\Big),
\label{D4pq}\\
g_3\big(\sigma(p_1,p_2,p_3,p_4)\big)=g_3\big(q_1=p_{\sigma(1)},q_2=p_{\sigma(2)},q_3=p_{\sigma(3)},q_4=p_{\sigma(4)}\big).
\label{g3pq}
\end{gather}
The composite $\delta$ function $\delta\Big(D_4\big(\sigma(p_1,p_2,p_3,p_4)\big)\Big)$ is then evaluated with respect to $p_4$:\footnote{We find necessary to evaluate the composite $\delta$ functions during the formulation of correlation functions because in loop calculation the loop momenta on the vertex should stay fixed (for example in a tadpole diagram). All we can generate through the composite $\delta$ function(s) is then how a certain external momentum becomes dependent on the other/others.}
\begin{equation}
\delta\Big(D_4\big(\sigma(p_1,p_2,p_3,p_4)\big)\Big)=\frac{\delta\Big(p_4-p_4(p_1,p_2,p_3)\Big)}{\det\left(\frac{\partial D_{4_\mu}\big(\sigma(p_1,p_2,p_3,p_4)\big)}{\partial p_{4_\nu}}\right)\bigg|_{p_4=p_4(p_1,p_2,p_3)}},
\label{deltaD4}
\end{equation}
where $p_4(p_1,p_2,p_3)$ is the solution to the equation
\begin{equation}
D_4\big(\sigma(p_1,p_2,p_3,p_4)\big)=0.
\label{DeltaD4=0}
\end{equation}
At $\beta^1$ order, this equation can be solved iteratively, noting that
\begin{equation}
D_4\big(\sigma(p_1,p_2,p_3,p_4)\big)=p_1+p_2+p_3+p_4+\beta D_4^1\big(\sigma(p_1,p_2,p_3,p_4)\big)+\mathcal O(\beta^2),
\label{D41234}
\end{equation}
thus the iterative solution of $p_4$ takes the following form:
\begin{equation}
\begin{split}
p_4(p_1,p_2,p_3)&=p_4^0(p_1,p_2,p_3)+\beta p_4^1(p_1,p_2,p_3)+\mathcal O(\beta^2)
\\&=-p_1-p_2-p_3-\beta D_4^1\big(\sigma(p_1,p_2,p_3,p_4^0=-p_1-p_2-p_3)\big)+\mathcal O(\beta^2).
\label{p4}
\end{split}
\end{equation}
Similarly, in order to obtain $G_{tree}(x_1,x_2,x_3,x_4)$ up to $\beta^1$ order, we have to expand the $g_3$ factor and the Jacobian determinant in \eqref{deltaD4} up to first order in $\beta$ around the  solution $p_4(p_1,p_2,p_3)$. This is straightforward since both of them have a constant value $1$ at $\beta^0$ order, and hence the expansion involves only expansions of these two objects up to $\beta^1$ order at the place $p_4=p_4^0=-p_1-p_2-p_3$. Moreover, at $\beta^1$ order, the determinant reduces to
\begin{equation}
\begin{split}
&\det\left(\frac{\partial D_{4_\mu}\big(\sigma(p_1,p_2,p_3,p_4)\big)}{\partial p_{4_\nu}}\right)\bigg|_{p_4=p_4^0}
=1+{\rm tr}\frac{\partial D^1_{4_\mu}\big(\sigma(p_1,p_2,p_3,p_4)\big)}{\partial p_{4_\nu}}\bigg|_{p_4=p_4^0}+\mathcal O(\beta^2).
\label{detD4mu}
\end{split}
\end{equation}
Finally, we also notice that the momentum in the last external propagator is shifted from the commutative solution $p_4^0$. We, therefore, expand it to $\beta^1$ order, too, obtaining
\begin{equation}
\begin{split}
\frac{e^{ip_4(p_1,p_2,p_3)x_4}}{p_4(p_1,p_2,p_3)^2+m^2}=&\frac{e^{-i(p_1+p_2+p_3)x_4}}{(p_1+p_2+p_3)^2+m^2}
\\&\cdot\left(1+\beta p_4^1(p_1,p_2,p_3)\cdot\Big(ix_4+\frac{2(p_1+p_2+p_3)}{(p_1+p_2+p_3)^2+m^2}\Big)\right)+\mathcal O(\beta^2).
\label{ep4x4}
\end{split}
\end{equation}

Now we collect all $\beta^1$-order contributions and sum over the $S_4$ permutations to obtain
\begin{equation}
\begin{split}
&G_{tree}(x_1,x_2,x_3,x_4)=G_{tree}^0\left(x_1,x_2,x_3,x_4\right)+\beta G_{tree}^1\left(x_1,x_2,x_3,x_4\right)+\mathcal O(\beta^2)
\\=&\lambda\int\frac{d^4 p_1}{(2\pi)^4}\frac{d^4 p_2}{(2\pi)^4}\frac{d^4 p_3}{(2\pi)^4}
\frac{e^{ip_1x_1}}{p_1^2+m^2}\frac{e^{ip_2x_2}}{p_2^2+m^2}\frac{e^{ip_3x_3}}{p_3^2+m^2}\frac{e^{-i(p_1+p_2+p_3)x_4}}{(p_1+p_2+p_3)^2+m^2}
\\&\cdot\Bigg(1+\frac{\beta}{3}\bigg(\Sigma_1+\Sigma_2\cdot\Big(ix_4+\frac{2(p_1+p_2+p_3)}{(p_1+p_2+p_3)^2+m^2}\Big)\bigg)\Bigg)+\mathcal O(\beta^2),
\end{split}
\label{Gtree1234}
\end{equation}
where
\begin{gather}
\Sigma_1(p_1,p_2,p_3)=(D+2)(s_1+s_2)\big(p_1\cdot p_2+p_2\cdot p_3+p_3\cdot p_1\big),
\label{Sigma1}\\
\begin{split}
\Sigma_2(p_1,p_2,p_3)=&-(s_1+s_2)\Big(p_1\big((p_1+p_2+p_3)^2-p_1^2\big)
\\&+p_2\big((p_1+p_2+p_3)^2-p_2^2\big)+p_3\big((p_1+p_2+p_3)^2-p_3^2\big)\Big).
\label{Sigma2}
\end{split}
\end{gather}

\subsection{One-loop two-point function}
\begin{figure}[t]
\begin{center}
\includegraphics[width=5cm,angle=0]{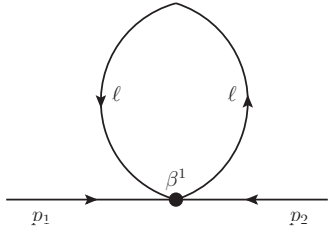}
\end{center}
\caption{Tadpole contribution to the two-point function.}
\label{fig:FD2}
\end{figure}
Following the same procedure as for the tree-level four-point function, we can now evaluate the one-loop two-point function of Fig.\ref{fig:FD2},
\begin{equation}
\begin{split}
G_{1-loop}(x_1,x_2)=&\frac{1}{2}\frac{\lambda}{4!}\int\frac{d^D p_1}{(2\pi)^D}\frac{d^D p_2}{(2\pi)^D}\frac{d^D \ell}{(2\pi)^D}\frac{e^{ip_1x_1}}{p_1^2+m^2}\frac{e^{ip_2x_2}}{p_2^2+m^2}\frac{1}{\ell^2+m^2}
\\&\cdot(2\pi)^4\sum\limits_{\sigma\in S_4}\delta\Big(D_4\big(\sigma(p_1,p_2,\ell,-\ell)\big)\Big)\cdot g_3\Big(\sigma\big(p_1,p_2,\ell,-\ell)\big)\Big).
\label{G1-loopDg}
\end{split}
\end{equation}
A peculiar property of Snyder and some other noncommutative field theories \cite{AmelinoCamelia:2001fd} is that, due to the law of addition of the momenta, $p_1$ and $p_2$ are, in general, different, so the momenta are not strictly conserved due to loop effects.

All $\delta$ functions in \eqref{G1-loopDg} can be evaluated using the iterative procedure of subsection IVA. After summing over all these permutation channels, we observe that the structures $\Sigma_1$ and $\Sigma_2$ emerge as expected. Using $\Sigma_1$ and $\Sigma_2$, we can rewrite $G_{1-loop}(x_1,x_2)$ as follows
\begin{equation}
\begin{split}
G_{1-loop}(x_1,x_2)=&\frac{\lambda}{2}\int\frac{d^D p_1}{(2\pi)^D}\frac{e^{ip_1x_1}}{p_1^2+m^2}\frac{e^{-ip_1x_2}}{p_1^2+m^2}
\int\frac{d^D \ell}{(2\pi)^D}\frac{1}{\ell^2+m^2}\Bigg(1+\frac{\beta}{3}\bigg(\Sigma_1(p_1,\ell,-\ell)
\\&\quad\quad\quad+\Sigma_2(p_1,\ell,-\ell)\cdot\Big(ix_4+\frac{2p_1}{p_1^2+m^2}\Big)\bigg)\Bigg)+\mathcal O(\beta^2).
\end{split}
\label{G1-loopbeta1}
\end{equation}
Once we evaluate $\Sigma_1$ and $\Sigma_2$ explicitly, an intriguing cancellation happens to send $\Sigma_2$ to zero and erases the effect of momentum nonconservation completely. The one-loop, two-point function then boils down to
\begin{equation}
\begin{split}
G_{1-loop}(x_1,x_2)=&\frac{\lambda}{2}\int\frac{d^D p_1}{(2\pi)^D}\frac{e^{ip_1x_1}}{p_1^2+m^2}\frac{e^{-ip_1x_2}}{p_1^2+m^2}
\int\frac{d^D \ell}{(2\pi)^D}\frac{1}{\ell^2+m^2}\left(1-\frac{\beta}{3}(D+2)(s_1+s_2)\ell^2\right)+\mathcal O(\beta^2)
\\&=\frac{\lambda}{2}\int\frac{d^D p_1}{(2\pi)^D}\frac{e^{ip_1x_1}}{p_1^2+m^2}\frac{e^{-ip_1x_2}}{p_1^2+m^2}
\left(1+m^2\frac{\beta}{3}(D+2)(s_1+s_2)\right)(4\pi)^{-\frac{D}{2}}m^{D-2}\Gamma\left(1-\frac{D}{2}\right).
\label{G1-loopbeta2}
\end{split}
\end{equation}
While the integral is quartic divergent, the Green function has the same structure as at tree level, thus one could, in principle, renormalize it using a mass counter-term $\delta m^2$.

\subsection{One-loop, four-point function}
\begin{figure}[t]
\begin{center}
\includegraphics[width=10cm,angle=0]{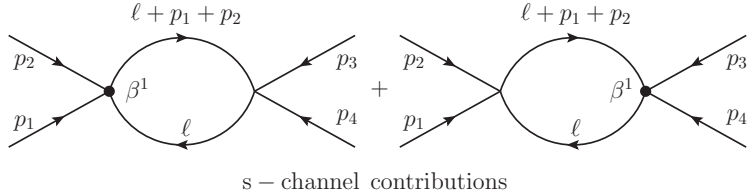}
\end{center}
\caption{Bubble diagram contributions to the four-point function in the s channel.}
\label{fig:FD3}
\end{figure}

\begin{figure}[t]
\begin{center}
\includegraphics[width=10cm,angle=0]{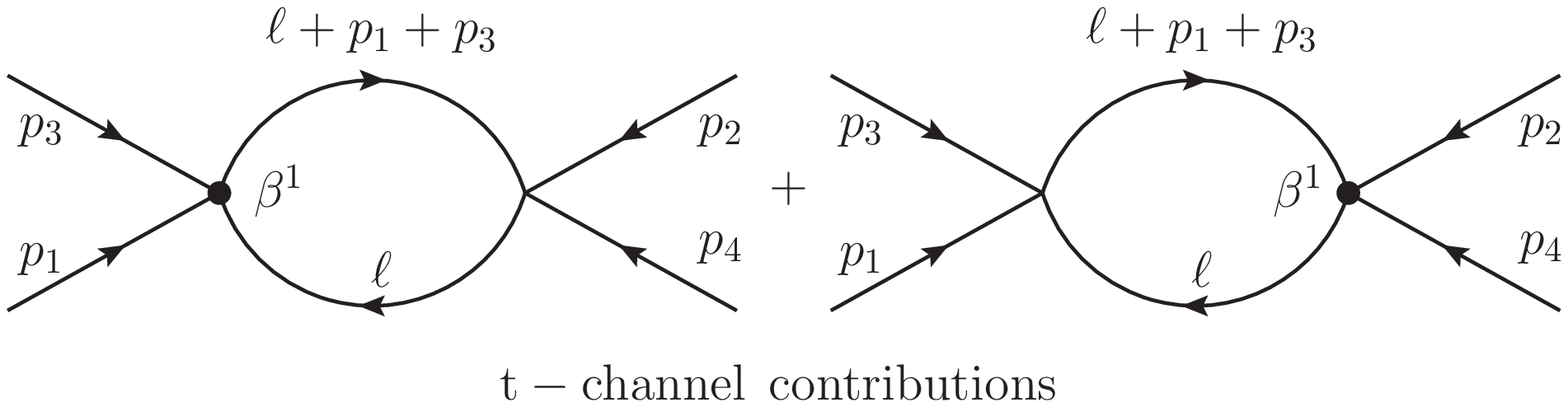}
\end{center}
\caption{Bubble diagram contributions to the four-point function in the t channel.}
\label{fig:FD4}
\end{figure}

\begin{figure}[t]
\begin{center}
\includegraphics[width=10cm,angle=0]{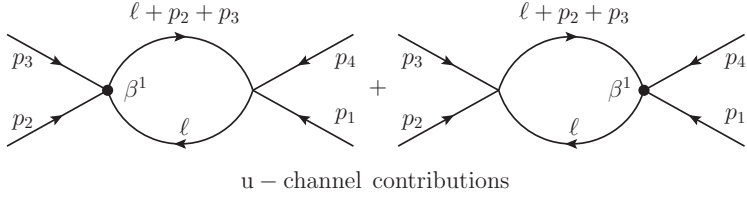}
\end{center}
\caption{Bubble diagram contributions to the four-point function in the u channel.}
\label{fig:FD5}
\end{figure}

As the commutative counterpart, one-loop four-point function can still be split into three Mandelstam-variable channels, as depicted in Figs. \ref{fig:FD3}--\ref{fig:FD5}
 \begin{equation}
G_{1-loop}(x_1,x_2,x_3,x_4)=I_s+I_t+I_u,
\label{G1-loopstu}
\end{equation}
but each of them now splits into two, depending on which of the two vertices is evaluated to the $\beta^1$ order, as we choose once again to integrate over the external momentum $p_4$ only.
Note that this procedure creates an additional momentum shift within the loop-integral when $p_4$ is attached to the $\beta^0$ vertex which is not explicitly shown in the diagrams. By realizing that the $\beta^0$ vertex is totally symmetric with respect to all momenta attached, we are able, from Fig.\ref{fig:FD3}, to obtain the following expression
\begin{equation}
I_s=I_s^0+\beta(I_1+I_2)+\mathcal O(\beta^2),
\label{Is}
\end{equation}
with
\begin{equation}
\begin{split}
I_s^0=&\frac{\lambda^2}{2}\int\frac{d^D p_1}{(2\pi)^D}\frac{d^D p_2}{(2\pi)^D}\frac{d^D p_3}{(2\pi)^D}\frac{e^{ip_1x_1}}{p_1^2+m^2}\frac{e^{ip_2x_2}}{p_2^2+m^2}\frac{e^{ip_3x_3}}{p_3^2+m^2}\frac{e^{-i(p_1+p_2+p_3)x_4}}{(p_1+p_2+p_3)^2+m^2}
\\&\cdot\int\frac{d^D \ell}{(2\pi)^D}\frac{1}{(\ell^2+m^2)((\ell+p_1+p_2)^2+m^2)},
\end{split}
\label{Is0}
\end{equation}
the usual $\beta^0$-order loop contribution, while
\begin{equation}
\begin{split}
I_1=&\frac{\lambda^2}{6}\int\frac{d^D p_1}{(2\pi)^D}\frac{d^D p_2}{(2\pi)^D}\frac{d^D p_3}{(2\pi)^D}
\frac{e^{ip_1x_1}}{p_1^2+m^2}\frac{e^{ip_2x_2}}{p_2^2+m^2}\frac{e^{ip_3x_3}}{p_3^2+m^2}\frac{e^{-i(p_1+p_2+p_3)x_4}}{(p_1+p_2+p_3)^2+m^2}
\\&\cdot\int\frac{d^D \ell}{(2\pi)^D}\frac{1}{(\ell^2+m^2)((\ell+p_1+p_2)^2+m^2)}
\\&\cdot\Bigg(\Sigma_1(-\ell,\ell+p_1+p_2,p_3)
+\Sigma_2(-\ell,\ell+p_1+p_2,p_3)\bigg(ix_4+\frac{2(p_1+p_2+p_3)}{(p_1+p_2+p_3)^2+m^2}\bigg)\Bigg),
\end{split}
\label{I1}
\end{equation}
and
\begin{equation}
\begin{split}
I_2=&\frac{\lambda^2}{6}\int\frac{d^D p_1}{(2\pi)^D}\frac{d^D p_2}{(2\pi)^D}\frac{d^D p_3}{(2\pi)^D}
\frac{e^{ip_1x_1}}{p_1^2+m^2}\frac{e^{ip_2x_2}}{p_2^2+m^2}\frac{e^{ip_3x_3}}{p_3^2+m^2}\frac{e^{-i(p_1+p_2+p_3)x_4}}{(p_1+p_2+p_3)^2+m^2}
\\&\cdot\int\frac{d^D \ell}{(2\pi)^D}\frac{1}{(\ell^2+m^2)((\ell+p_1+p_2)^2+m^2)}
\\&\cdot\Bigg(\Sigma_1(p_1,p_2,\ell)
+\Sigma_2(p_1,p_2,\ell)\bigg(ix_4+\frac{2(p_1+p_2+p_3)}{(p_1+p_2+p_3)^2+m^2}+\frac{2(\ell+p_1+p_2)}{(\ell+p_1+p_2)^2+m^2}\bigg)\Bigg),
\end{split}
\label{I2}
\end{equation}
are the $\beta^1$-order corrections from Snyder-type deformations. Once we work out all the objects explicitly, the s-channel integral boils down to
\begin{equation}
I_s=G_{tree}^0(x_1,x_2,x_3,x_4)\cdot\big(\mathcal I_s+\beta(\mathcal I_1+\mathcal I_2)\big)+\beta G_{tree}^1(x_1,x_2,x_3,x_4)\cdot\mathcal I_s,
\label{4pointstructure}
\end{equation}
where
\begin{equation}
\mathcal I_s=\int\frac{d^D \ell}{(2\pi)^D}\frac{\lambda}{(\ell^2+m^2)((\ell+p_1+p_2)^2+m^2)},
\label{calIs}
\end{equation}
is the usual s-channel scalar loop integral
while
\begin{equation}
\begin{split}
\mathcal I_1=&-\frac{\lambda}{6}
\int\frac{d^D \ell}{(2\pi)^D}\frac{(D+2)(s_1+s_2)\ell^2}{(\ell^2+m^2)((\ell+p_1+p_2)^2+m^2)},
\label{calI1}
\end{split}
\end{equation}
and
\begin{equation}
\begin{split}
\mathcal I_2=&\frac{\lambda}{6}
\int\frac{d^D \ell}{(2\pi)^D}\frac{2(\ell+p_1+p_2)\Sigma_2(p_1,p_2,\ell)}{(\ell^2+m^2)((\ell+p_1+p_2)^2+m^2)^2},
\end{split}
\label{calI2}
\end{equation}
presents Snyder-type deformation effects within the loop integral at $\beta^1$ order. We are particularly interested in the UV divergence within these two integrals. It is easy to see that $\mathcal I_2$ is quadratic UV divergent. An explicit computation shows that in the $D\to 4-\epsilon$ limit this integral reduces to
\begin{equation}
\begin{split}
\mathcal I_1=&\lambda(s_1+s_2)\frac{m^2}{(4\pi)^2}\left(\frac{4}{\epsilon}+\frac{1}{3}-2\gamma_E
-\int\limits_0^1 dz\,\log \frac{m^2(z(1-z)(p_1+p_2)^2+m^2)}{(4\pi)^2}\right)+\mathcal O(\epsilon).
\end{split}
\label{calI1fin}
\end{equation}
The integral $\mathcal I_2$ requires a more detailed investigation. Writing down explicitly the numerator
\begin{equation}
\begin{split}
\mathcal I_2=&\frac{\lambda}{3}(s_1+s_2)
\int\frac{d^D \ell}{(2\pi)^D}\frac{(\ell+p_1+p_2)^4-(\ell+p_1+p_2)\cdot(p_1p_1^2+p_2p_2^2+\ell\ell^2)}{(\ell^2+m^2)((\ell+p_1+p_2)^2+m^2)^2}
\\=&\frac{\lambda}{3}(s_1+s_2)\int\limits_0^1 dz\,\int\frac{d^D \ell}{(2\pi)^D}\frac{2z(3z-2)(1+\frac{2}{D})(p_1+p_2)^2}{(\ell^2+z(1-z)(p_1+p_2)^2+m^2)^2}+{\rm finite\; terms},
\end{split}
\label{calI2fin}
\end{equation}
where $z$ is the usual Feynman variable. In the the $D\to 4-\epsilon$ limit {\bf the} integral reduces to
\begin{equation}
\begin{split}
\mathcal I_2=&\frac{\lambda}{3}(s_1+s_2)\frac{1}{(4\pi)^2}\left(1+\frac{1}{2}+\frac{\epsilon}{8}\right)(p_1+p_2)^2
\int\limits_0^1 dz\,2z(3z-2)
\\&\cdot\left(\frac{2}{\epsilon}-\gamma_E+\log4\pi-\log\Big(z(1-z)(p_1+p_2)^2+m^2\Big)
+\mathcal O(\epsilon)\right)
+{\rm finite\; terms}.
\end{split}
\label{calI2final}
\end{equation}
We can then find that the $1/\epsilon$ divergence vanishes because
\begin{equation}
\int\limits_0^1 2z(3z-2)=2(z^3-z^2)\Big|_0^1=0,
\label{Int0}
\end{equation}
therefore, the whole integral remains finite in dimensional regularization.

The t and u channels, corresponding to Figs.\ \ref{fig:FD4} and \ref{fig:FD5}, can be obtained from the s-channel formulas above by the permutations $p_2\leftrightarrow p_3$ and $p_1\leftrightarrow p_3$, respectively.

The one-loop structure~\eqref{4pointstructure} suggests that we should  renormalize the four-point function by introducing a $\beta$-expansion of the coupling constant counter term
\begin{equation}
\delta\lambda=\delta\lambda^0+\beta\delta\lambda^1+\mathcal O(\beta^2).
\label{dlambda}
\end{equation}
We see that the UV divergence in $\mathcal I_s$ can be absorbed by $\delta\lambda^0$, and the new divergence from $I_1$ by $\delta\lambda^1$. The latter is valid since the $1/\epsilon$ term is proportional to the mass only.

\subsection{UV divergence in the one-loop, six-point function}

\begin{figure}[t]
\begin{center}
\includegraphics[width=7cm,angle=0]{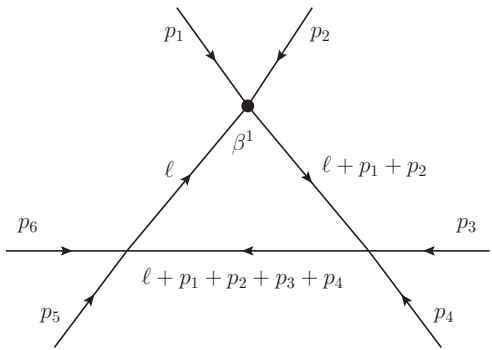}
\end{center}
\caption{Typical diagram contribution to the six-point function. The $\beta^1$-order contribution has to be considered as running over all three vertices in order to complete each channel.}
\label{fig:FD6}
\end{figure}

Our experience with two- and four-point function shows that the degree of divergence of each of them is higher than its commutative counterpart, which suggests that the one-loop, six-point function can pick up UV divergent contributions also from the triangle diagram of Fig.\ref{fig:FD6}, where the black dot represents the $\beta^1$ vertex which contains the $\Sigma_1(p_1,p_2,\ell)$ term. Explicit evaluation, starting from (\ref{Sint1}), gives the following form of the divergent integral in one channel:
\begin{equation}
\begin{split}
\mathcal I_6^{\rm UV}=&\int\frac{d^D\ell}{(2\pi)^D}\frac{\Sigma_1(p_1,p_2,\ell)+\Sigma_1(\ell+p_1+p_2,p_3,p_4)+\Sigma_1(\ell+p_1+p_2+p_3+p_4,p_5,-\ell)}{(\ell^2+m^2)((\ell+p_1+p_2)^2+m^2)((\ell+p_1+p_2+p_3+p_4)^2+m^2)}
\\
=&\int\frac{d^D\ell}{(2\pi)^D}\frac{(D+2)(s_1+s_2)\Big(-\ell^2+\sum\limits_{1\le i<j\le 5} p_i\cdot p_j\Big)}{(\ell^2+m^2)((\ell+p_1+p_2)^2+m^2)((\ell+p_1+p_2+p_3+p_4)^2+m^2)},
\end{split}
\label{I6}
\end{equation}
with $\Sigma_1$ being defined in (\ref{Sigma1}). The sum of three $\Sigma_1$'s contains also contributions from two additional diagrams obtained from the diagram in Fig.\ \ref{fig:FD6} by shifting the black dot to the other two available positions in the diagram. Other channels can be obtained by an appropriate permutation of the external momenta. As we can see, the first term in the numerator gives rise to a logarithmic UV divergence. However, we can of course still remove this divergence by demanding $s_1+s_2=0$. In this case all nontrivial $\beta^1$-order quantum corrections are removed and we are dealing with exactly the same renormalization procedure as in the commutative theory.

\section{The effect of Snyder nonassociativity}

The Snyder-type star products discussed in Sec. II are, in general, nonassociative, except in the case $s_2=2s_1$, which means that the ordering of the products matters. Taking into account integration by parts, from (\ref{Sint1}) we obtain two additional types of $\phi^4$ interactions, giving altogether the following:
\begin{eqnarray}
S_{int}^1\equiv(S_{int}^1)_1&=&-\frac{\lambda}{4!}\int \phi\big(\phi\star(\phi\star\phi)\big),
\label{AS1}\\
(S_{int}^1)_2&=&-\frac{\lambda}{4!}\int \phi\big((\phi\star\phi)\star\phi\big),
\label{AS2}\\
(S_{int}^1)_3&=&-\frac{\lambda}{4!}\int (\phi\star\phi)(\phi\star\phi).
\label{AS3}
\end{eqnarray}
Repeating the computation in prior sections, using $(S_{int}^1)_2$ and $(S_{int}^1)_3$ in place of $(S_{int}^1)_1$, we find that all three variants of the Snyder-type $\phi^4$ interaction give the same results at the first order in $\beta$. This result is rather surprising. Each of the permutation channels contains different inputs, yet the average over all permutations totally cancels all these effects. It is, however, possible that going to higher orders in $\beta$, this degeneracy is lost.

\section{On the Snyder-type realization with $s_1=-s_2$}

A particularly interesting result of our tree- and one-loop level study is that one special combination $s_1=-s_2$ removes all $\beta^1$-order corrections. As we are going to show below, it turns out that this point contains peculiar information also from the point of view of realizations.

A fundamental quantity in the realization approach  to the noncommutative space is the action of the NC wave operator on identity:
\begin{equation}
e^{i(k\hat x)}\triangleright 1=e^{iK(k)\cdot x} e^{iF(k)}
\label{58}
\end{equation}
For a general NC coordinate $\hat x^\mu=x^\alpha\varphi_\alpha^\mu(p)+\chi^\mu(p)$, $K^\mu(k)$ and $F(x)$ satisfy the following differential equations
\begin{gather}
\frac{dK^\mu(\lambda k)}{d\lambda}=k^\alpha\varphi_\alpha^\mu\left(K^\mu(\lambda k)\right),
\label{59}\\
\frac{dF(\lambda k)}{d\lambda}=k_\alpha\chi^\alpha\left(K^\mu(\lambda k)\right).
\label{60}
\end{gather}
For Snyder-type spaces it is natural to assume that $K^\mu(k)=k^\mu\mathcal{K}(\beta k^2)$, since there is no relevant tensor structure other than the Lorentz/Euclidean metric. Now for the Snyder-type realization $\varphi^\mu_\alpha(p)=\delta^\mu_\alpha(1+\beta s_1 p^2)+\beta s_2  p_\alpha p^\mu$, we have
\begin{equation}
\begin{split}
\frac{dK^\mu(\lambda k)}{d\lambda}&=k^\mu\mathcal K(\beta\lambda^2 k^2)+\lambda k^\mu\frac{d\mathcal K(\beta\lambda^2 k^2)}{d\lambda}
\\&=k^\alpha\Big(\delta^\mu_\alpha\big(1+\beta s_1 K^2(k)\big)+\beta s_2 K_\alpha(k)K^\mu(k)\Big)
\\&=k^\mu\Big(1+\beta \lambda^2 k^2(s_1+s_2)\mathcal K^2(\beta\lambda^2 k^2)\Big).
\end{split}
\label{61}
\end{equation}
One can then easily see that $\mathcal K=1$ when $s_1+s_2=0$, i.e. $K^\mu=k^\mu$. Such a realization is called Weyl realization in the literature, see \cite{Meljanac:2015zel} and references therein.

Furthermore, for Hermitian realizations of the Snyder-type spaces we have
\begin{equation}
\chi^\mu(p)=-\frac{1}{2}\Big[x^\alpha,\varphi^\mu_\alpha(p)\Big]=-\frac{i}{2}\frac{\partial\varphi^\mu_\alpha(p)}{\partial p_\alpha}=-i\beta\left(s_1+\frac{D+1}{2}s_2\right)p^\mu.
\label{62}
\end{equation}
Then, for the Weyl realization where $s_1+s_2=0$ (\ref{62}) reduces to
\begin{equation}
\chi^\mu(p)=-\frac{i}{2}\beta s_2(D-1)p^\mu.
\label{63}
\end{equation}
Therefore (\ref{60}) reads
\begin{equation}
\frac{dF(\lambda k)}{d\lambda}=-\frac{i}{2}\beta s_2\lambda(D-1)k^2,
\label{64}
\end{equation}
and its solution for $\lambda=1$ is given by
\begin{equation}
F(k)=-\frac{i}{4}\beta s_2(D-1)k^2.
\label{65}
\end{equation}

Finally, the fundamental relation between the product of two plane wave operators
\begin{equation}
e^{i(k\hat x)}\triangleright e^{i(q\hat x)}=e^{iP(k,q)x}e^{iQ(k,q)}
\label{66}
\end{equation}
and the star product of two plane wave functions
\begin{equation}
e^{i(k x)}\star e^{i(q x)}=e^{iD(k,q)x}e^{iG(k,q)}
\label{67}
\end{equation}
is also slightly simplified, namely
\begin{equation}
D(k,q)=P(k,q),\quad G(k,q)=Q(k,q)-F(k),
\label{68}
\end{equation}
since $K^\mu(k)$ is now trivial.

It remains to solve for $P^{\mu}(k,q)$ and $Q(k,q)$ completely. The authors expect that such solution can be found in the near future.

\section{Discussion and Conclusion}

In this article we have studied Snyder field theory with the action truncated at first order in the deformation parameter $\beta$, producing an effective model on commutative spacetime. The study is performed by using the functional method in momentum space up to one loop.

We recall the main points of our analysis: we have proposed a simple perturbative quantization for the $\phi^4$ theory on Snyder-type spaces with Hermitian realizations and have evaluated the one-loop, two- and four-point functions at $\beta^1$ order, showing that they give raise to UV divergences. They are stronger than in the commutative theory, but nevertheless they can be absorbed by the tree level counter-terms.

However, the $\beta^1$ order one-loop, six-point function receives a logarithmic UV divergent quantum correction in general, which renders the theory unrenormalizable. Remarkably, at $\beta^1$ order all information about nonassociativity in the definition of $\phi^4$ interaction is canceled, namely one obtains identical results for both the tree and the one-loop correlation functions independently of the ordering of the products.

Inspecting the $\beta^1$-order equations (\ref{Sigma1}), (\ref{Sigma2}), (\ref{calI1fin}), (\ref{calI2fin}), (\ref{I6}) we find that the correlation functions depend on the free parameters $s_1$ and $s_2$ only through their sum $s_1+s_2$. In other words, one can turn off all nontrivial $\beta^1$-order effects by setting $s_1=-s_2$, which corresponds to the removal of the dependence on the dilatation operator $(x\cdot p)$ from the definition of the noncommutative coordinates $\hat x_\mu$ in \eqref{xreal}.

Generally speaking, the effects of noncommutativity can only be properly displayed when the star product is treated nonperturbatively, since any truncation up to a certain order of the deformation parameter would normally remove nontrivial effects. However, certain special cancellations of divergences found after the truncation may remain partially valid in the full theory~\cite{Horvat:2013rga}. From this perspective the special features of the point $s_1=-s_2$ found in this work could maintain their importance. In fact, this special point does lead to certain nontrivial $\beta$-exact structure in the determination of realizations, as shown in Sec. VI.

As already mentioned above, so far our investigation has been limited to the first order in the $\beta$-deformation parameter. The full theory has of course different properties, especially in the UV limit, which could be finite for some choices of the defining commutation relations. For example, let us consider the case of the original Snyder model \cite{Snyder:1946qz} corresponding to $s_1=0$, $s_2=1$: in the full theory the cyclicity condition still holds, so that the propagators are the same as in the linearized theory, while the vertices take the form
\begin{equation}
G^{tree}(p_1,p_2,p_3,p_4)={\lambda\over4!}\ {\delta\big(D_4(p_1,p_2,p_3,p_4)\big)\over\big[(1+\beta p_2\cdot D(p_3,p_4))(1+\beta p_3\cdot p_4)\big]^ {5/2}}.
\end{equation}
The extra terms in the denominator with respect to the commutative case improve notably the convergence properties of the loop integrals in the UV regime, and would likely render them finite. This should, however, be checked explicitly. It is also possible that the problems due to non-conservation of momenta in loops are solved as in the linearized theory, when the average over the ordering of the lines entering a vertex is performed.

A rigorous proof of these properties is, obviously, difficult, since the calculations are rather involved. This problem is currently under study. Our general formalism for the generating functional may be a good starting point towards an investigation of the full theory. We hope that the special cancellation point $s_1=-s_2$ can be revisited and play a role within the framework of the full theory too.

\section{Acknowledgements}
This work is supported by the Croatian Science Foundation (HRZZ)
under Contract No. IP-2014-09-9582. We acknowledge the support of the COST Action MP1405  (QSPACE).
S.Meljanac and J.You acknowledge support by the H2020 Twining project No. 692194, RBI-T-WINNING. J.Trampetic and J.You would like to acknowledge the support of W. Hollik and the Max-Planck-Institute for Physics, Munich, for hospitality.
A great deal of computation was done by using $\rm MATHEMATICA$ {\bf 8.0} \cite{mathematica} plus the tensor algebra package xACT~\cite{xAct}. Special thanks to A. Ilakovac and D. Kekez for the computer software/hardware support.


\begin{thebibliography}{999}

\bibitem{Seiberg:1999vs}
  N.~Seiberg and E.~Witten,
J. High Energy Phys. {\bf 09} (1999) 032.  

\bibitem{Connes:1994yd}
  A.~Connes, 
  {\it Noncommutative Geometry} (Academic  Press, New York, 1994).

\bibitem{Doplicher:1994tu}
  S.~Doplicher, K.~Fredenhagen, and J.~E.~Roberts,
  Commun.\ Math.\ Phys.\  {\bf 172}, 187 (1995) 

\bibitem{Majid:1996kd}
  S.~Majid,
 {\it Foundations of Quantum Group Theory},
 (Cambridge University Press, Cambridge, England, 1995).

\bibitem{Landi:1997sh}
  G.~Landi,
   {\it Introduction to Noncommutative Spaces and Their Geometry,}
  (Springer, New York, 1997),
  Lect.\ Notes Phys.\ Monogr.\  {\bf 51}, 1 (1997).

\bibitem{Madore:1999bi}
  J.~Madore,
 {\it Noncommutative Geometry for Pedestrians}, 
 arXiv:gr-qc/9906059.

\bibitem{Madore:2000aq}
  J.~Madore,
 {\it An Introduction to Noncommutative Differential Geometry and its Physical Applications},
  Lond.\ Math.\ Soc.\ Lect.\ Note Ser.\  {\bf 257}, 1 (2000).
 

\bibitem{GraciaBondia:2001tr}
  J.~M.~Gracia-Bondia, J.~C.~Varilly, and H.~Figueroa,
  {\it  Elements Of Noncommutative Geometry},
   (Birkhaeuser, Boston, 2001).

\bibitem{Szabo:2001kg}
  R.~J.~Szabo,
  Phys.\ Rep.\  {\bf 378}, 207 (2003).  

\bibitem{Szabo:2009tn}
  R.~J.~Szabo,
  Gen.\ Relativ.\ Gravit.\  {\bf 42 }, 1 (2010). 


\bibitem{Moyal:1949sk}
  J.~E.~Moyal,
  Proc.\ Cambridge Philos.\ Soc.\  {\bf 45}, 99 (1949).  

\bibitem{Douglas:2001ba}
  M.~R.~Douglas and N.~A.~Nekrasov,
  Rev.\ Mod.\ Phys.\  {\bf 73}, 977 (2001).

\bibitem{Schupp:2002up}
  P.~Schupp, J.~Trampetic, J.~Wess, and G.~Raffelt,
  Eur.\ Phys.\ J.\ C {\bf 36}, 405 (2004).


\bibitem{Schupp:2008fs}
  P.~Schupp and J.~You,
  J. High Energy Phys.  {\bf 08} (2008) 107.
  
\bibitem{Aharony:2000gz} 
  O.~Aharony, J.~Gomis, and T.~Mehen,
   J. High Energy Phys.  {\bf 09}, (2000) 023.

\bibitem{Horvat:2011bs}
  R.~Horvat, A.~Ilakovac, J.~Trampetic, and J.~You,
  J. High Energy Phys.  {\bf 12} (2011) 081.  



\bibitem{Martin:2016zon}
  C.~P.~Martin, J.~Trampetic and J.~You,
  J. High Energy Phys.  {\bf 05} (2016) 169.

\bibitem{Martin:2016hji}
  C.~P.~Martin, J.~Trampetic, and J.~You,
  Phys.\ Rev.\ D {\bf 94}, 041703 (2016). 
  
\bibitem{Martin:2016saw} 
  C.~P.~Martin, J.~Trampetic, and J.~You,
   J. High Energy Phys.  {\bf 09}, (2016) 052.

\bibitem{Horvat:2017gfm}
  R.~Horvat, J.~Trampetic, and J.~You,
  Phys.\ Lett.\ B {\bf 772}, 130 (2017).

\bibitem{Lukierski:1991pn}
  J.~Lukierski, H.~Ruegg, A.~Nowicki, and V.~N.~Tolstoi,
  Phys.\ Lett.\  B {\bf 264}, 331 (1991).

\bibitem{Lukierski:1992dt}
  J.~Lukierski, A.~Nowicki, and H.~Ruegg,
  Phys.\ Lett.\  B {\bf 293}, 344 (1992).

\bibitem{Meljanac:2007xb}
  S.~Meljanac, A.~Samsarov, M.~Stojic, and K.~S.~Gupta,
  Eur.\ Phys.\ J.\ C {\bf 53}, 295 (2008).

\bibitem{Govindarajan:2008qa}
  T.~R.~Govindarajan, K.~S.~Gupta, E.~Harikumar, S.~Meljanac, and D.~Meljanac,
  Phys.\ Rev.\ D {\bf 77}, 105010 (2008).

\bibitem{Meljanac:2010qp}
  S.~Meljanac and S.~Kresic-Juric,
  Int.\ J.\ Mod.\ Phys.\ A {\bf 26}, 3385 (2011).

\bibitem{Meljanac:2011cs}
  S.~Meljanac, A.~Samsarov, J.~Trampetic, and M.~Wohlgenannt,
   J. High Energy Phys.  {\bf 12} (2011) 010.

\bibitem{Govindarajan:2009wt} 
  T.~R.~Govindarajan, K.~S.~Gupta, E.~Harikumar, S.~Meljanac, and D.~Meljanac,
  Phys.\ Rev.\ D {\bf 80}, 025014 (2009).

\bibitem{Snyder:1946qz}
  H.~S.~Snyder,
  Phys.\ Rev.\  {\bf 71}, 38 (1947).

\bibitem{Maggiore:1993kv}
  M.~Maggiore,
 Phys.\ Lett.\ B {\bf 319}, 83 (1993).

\bibitem{Battisti:2010sr}
  M.~V.~Battisti and S.~Meljanac,
  Phys.\ Rev.\ D {\bf 82}, 024028 (2010).

\bibitem{Mignemi:2013aua}
  S.~Mignemi,
  Int.\ J.\ Mod.\ Phys.\ D {\bf 24}, 1550043 (2015). 

\bibitem{Mignemi:2015fva}
  S.~Mignemi and R.~Strajn,
  Phys.\ Lett.\ A {\bf 380}, 1714 (2016).

\bibitem{Lu:2011it}
  L.~Lu and A.~Stern,
  Nucl.\ Phys.\ B {\bf 854}, 894 (2012).

\bibitem{Lu:2011fh}
  L.~Lu and A.~Stern,
  Nucl.\ Phys.\ B {\bf 860}, 186 (2012).

\bibitem{Girelli:2010wi}
  F.~Girelli and E.~R.~Livine,
  J. High Energy Phys. {\bf 03} (2011) 132.

\bibitem{Meljanac:2016gbj}
  S.~Meljanac, D.~Meljanac, S.~Mignemi, and R.~\v Strajn,
arXiv:1608.06207 [hep-th].

\bibitem{Meljanac:2016jwk}
  S.~Meljanac, D.~Meljanac, F.~Mercati, and D.~Pikutic,
  Phys.\ Lett.\ B {\bf 766}, 181 (2017).

\bibitem{Meljanac:2017ikx}
  S.~Meljanac, D.~Meljanac, S.~Mignemi, and R.~\v Strajn,
 Phys.\ Lett.\ B {\bf 768}, 321 (2017).



\bibitem{Kupriyanov:2017oob}
  V.~G.~Kupriyanov and R.~J.~Szabo,
  J. High Energy Phys.  {\bf 02} (2017) 099. 


\bibitem{Minwalla:1999px}
  S.~Minwalla, M.~Van Raamsdonk, and N.~Seiberg,
  J. High Energy Phys.  {\bf 0002} (2000) 020.  
  
\bibitem{Grosse:2005iz} 
  H.~Grosse and M.~Wohlgenannt,
  Nucl.\ Phys.\ B {\bf 748}, 473 (2006).
  
\bibitem{Cohen:1998zx} 
  A.~G.~Cohen, D.~B.~Kaplan and A.~E.~Nelson,
  Phys.\ Rev.\ Lett.\  {\bf 82}, 4971 (1999).

\bibitem{Horvat:2010km} 
  R.~Horvat and J.~Trampetic,
  J. High Energy Phys.  {\bf 01}, (2011) 112.
 
\bibitem{AmelinoCamelia:2001fd}
 G.~Amelino-Camelia and M.~Arzano,
  Phys.\ Rev.\ D {\bf 65}, 084044 (2002).
  
\bibitem{Meljanac:2015zel} 
  S.~Meljanac, S.~Kresic-Juric and T.~Martinic,
  J.\ Math.\ Phys.\  {\bf 57}, 051704 (2016).

  
\bibitem{Horvat:2013rga} 
  R.~Horvat, A.~Ilakovac, J.~Trampetic and J.~You,
  J. High Energy Phys.  {\bf 11}, (2013) 071.

\bibitem{mathematica}
{\it Mathematica, Version 8.0} 
(Wolfram Research, Inc., Champaign, IL, 2010).

\bibitem{xAct}
J. Martin-Garcia, {\it xAct,} \href{http://www.xact.es/}
{http://www.xact.es/}.

\end{thebibliography}
\end{document}